\documentclass[preprint,12pt,authoryear]{elsarticle}

\usepackage{subcaption}
\usepackage{amssymb}
\usepackage{float}
\journal{Progress in Organic Coatings}

\begin{document}

\begin{frontmatter}



\title{On the theoretical framework for meniscus-guided manufacturing of large-area OPV modules}

\author[THN]{Fabian Gumpert}
\author[THN,FAU,HIERN]{Annika Janßen}
\author[FAU]{Robin Basu}
\author[FAU,HIERN]{Christoph J. Brabec}
\author[FAU,HIERN]{Hans-Joachim Egelhaaf}
\author[THN]{Jan Lohbreier}
\author[FAU]{Andreas Distler}

\affiliation[THN]{organization={Faculty of Applied Mathematics, Physics and Humanities, Nuremberg Institute of Technology},
            addressline={Keßlerplatz 12}, 
            city={Nuremberg},
            postcode={90489}, 
            state={Bavaria},
            country={Germany}}

\affiliation[FAU]{organization={Faculty of Engineering, Department of Material
Science, Materials for Electronics and Energy Technology (i-MEET), Friedrich-Alexander-Universität Erlangen-Nürnberg},
            addressline={Martensstraße 7}, 
            city={Erlangen},
            postcode={91058}, 
            state={Bavaria},
            country={Germany}}

\affiliation[HIERN]{organization={Helmholtz Institute Erlangen-Nürnberg for Renewable Energy (HI-ERN), Forschungszentrum Jülich GmbH},
            addressline={Immerwahrstraße 2}, 
            city={Erlangen},
            postcode={91058}, 
            state={Bavaria},
            country={Germany}}

\begin{abstract}
For the manufacturing of thin films of solution-processable organic semiconductors, e.g. for organic photovoltaics (OPV), meniscus guided-coating techniques are the method of choice for large-scale industrial applications. However, the process requires an in-depth understanding of the respective fluid dynamics to control the resulting film thickness. In this article, we derive an analytical expression to describe the layer thickness of coatings manufactured with a trapezoidal-shaped applicator as a function of various fluid and process parameters. The analytical calculations are compared with results from computational fluid dynamics (CFD) simulations and experimental data for an industrially relevant OPV active material system. The analytical calculations are compared with results from computational fluid dynamics (CFD) simulations and experimental data for an industrially relevant OPV active material system. The good agreement of all three approaches demonstrates the potential of the analytical and simulative methods to minimize the number of time- and resource-consuming experiments. Furthermore, our theoretical model can be used to enhance the homogeneity of large-area coatings by means of an acceleration profile of the applicator that can compensate the liquid loss during the coating process. The respective analytical expression is validated by simulated and experimentally obtained data for long-distance coatings. Finally, this approach is used to fabricate a large-area OPV module with new world record efficiency. 

\end{abstract}

\begin{keyword}
Organic photovoltaics \sep Meniscus-guided coating \sep Numerical simulation \sep Analytical description \sep Thin films \sep Accelerated coating velocity \sep Uniform film thickness \sep Organic semiconductor  \sep Computational fluid dynamics


\end{keyword}

\end{frontmatter}


\section{Introduction}\label{}
\begin{figure}[H]
    \centering
    \includegraphics[width=0.8\textwidth]{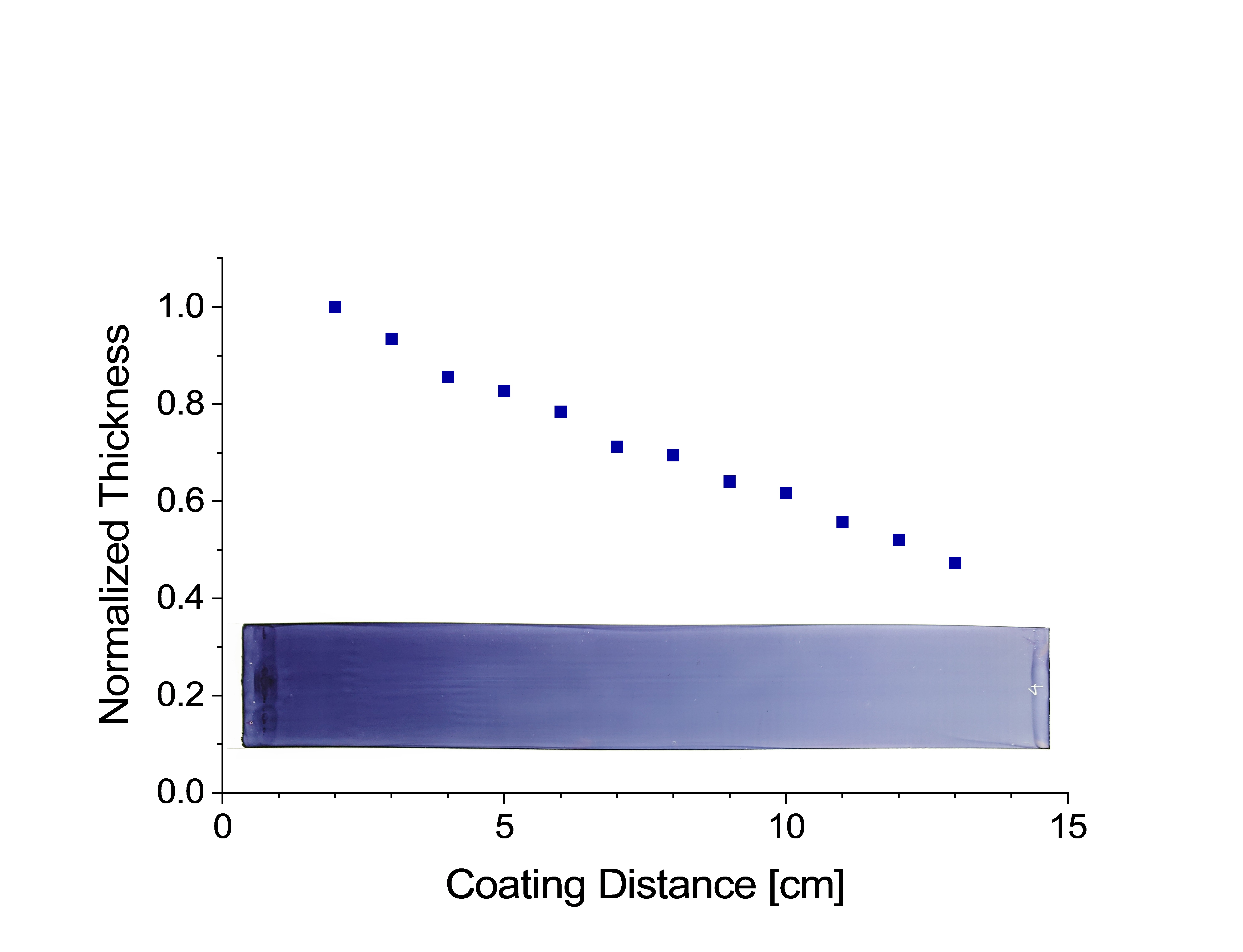}
    \caption{Normalized thickness profile of a P3HT:O-IDTBR film, which is manufactured by blade coating with constant velocity, over a coating distance of 15 $cm$. The steady decrease of film thickness with coating distance can be observed in the experimentally obtained data points and is also clearly visible in the photograph of the respective sample.}
    \label{figUneven}
\end{figure}

The importance of photovoltaics has grown in recent years as the demand for renewable energy continues to increase. Solution-processable organic photovoltaics (OPV) receive more and more scientific and economical attention due to their unique characteristics, like light weight, high throughput, and semi-transparency \cite{Hu2022}. The respective printing inks comprise the photoactive material dissolved in a solvent. After the deposition of the coating ink, the solvent evaporates during the drying process, leaving a dry film as the functional layer. Recent material developments result in photoelectric conversion efficiencies (PCEs) above 19 $\%$ \cite{Sun2022, Zhan2022, Zhu2022} for small-scale cells (few $mm^2$) at laboratory level. Uniform and even layers with a predefined thickness are required to achieve these record efficiencies.\\
Solution-processable techniques are preferred processes for industrial applications, since they  enable higher throughput (e.g. roll-to-roll manufacturing) and thus, lower costs. For roll-to-roll manufacturing, meniscus guided coating techniques are established methods to deposit the OPV materials \cite{NG2022, Gu2017, Berny2015}. Among these techniques, blade coating is scientifically and economically the most relevant technique, especially for the research on upscaling from small-scale cells to large-area modules. The upscaling to module size is still a major challenge that requires further investigation of the process \cite{Zhang2022}. Insufficient control and/or understanding of the deposition process results in uneven and non-uniform layer thicknesses and thus, to a decrease of the power conversion efficiency of the module.\\
In the blade coating process, an initial fluid volume is deposited between the applicator and the substrate to be coated. During the coating process, the applicator is moved horizontally and, by this, fluid is partially deposited onto the substrate. Consequently, the fluid volume beneath the applicator steadily decreases with coating distance. However, the fluid volume beneath the applicator influences the radius of the down-stream meniscus, which is essential for the deposited wet film thickness \cite{Gumpert_Janßen_Brabec_Egelhaaf_Lohbreier_Distler_2023}. As a consequence, the consumption of the fluid during the coating process leads to an uneven film thickness, i.e. an initially thicker film that becomes gradually thinner with coating distance, as shown in Figure \ref{figUneven}.\\

Besides laboratory experiments, numerical simulation and analytical description are established methods to better understand the formation of the wet film and to optimize the deposition process. Moreover, simulation and theory approaches further decrease the number of time- and resource-expensive experiments.
For example the influence of different blade geometries on the blade coating process are numerically investigated in \cite{Iliopoulos2005}. CFD models are reported to investigate the blade coating process as a function of different variables (e.g. substrate speed and fluid properties) \cite{Singh2021, Schmidt2009, mitsoulis2010numerical}. The theory of wet film formation is based on the work of Landau, Levich and Derjaguin \cite{Landau_Levich_1988, Derjaguin_1993}. Based on this theory, Gutenev et al. derived analytical expressions in \cite{Gutenev2003} to further describe the blade coating process. According to the theory of wet film formation, the coating velocity can compensate the fluid loss beneath the applicator. In \cite{Park_Han_2009, Tsai2015}, the influence of the coating velocity on the wet film thickness was experimentally investigated. In both articles, the theoretically predicted dependency of the film thickness on the coating velocity is confirmed.
In \cite{Gumpert_Janßen_Brabec_Egelhaaf_Lohbreier_Distler_2023}, we developed a numerical model of the coating process where the simulated thicknesses are in very good agreement with the experimental data. Furthermore, a linear function was derived which is able to predict the wet film thickness for a wide range of velocity and initial volume combinations.\\
In this article, we propose a complete theoretical framework for blade coating with a trapezoidal applicator shape, which does not only describe the influence of coating speed and ink volume, but also includes the ink properties, namely viscosity and surface tension. \\
In addition, for both the linear function of the previous work and the analytical formula, an expression is derived to describe an accelerated coating process to generate an even and uniform wet film with a constant targeted thickness. Finally, experiments are performed where the velocity of the applicator is increased according to the derived expressions. The resulting wet films provide a thickness as uniform as predicted and thus validate our simulations and analytical model. Thus, they prove to be powerful tools for all kinds of research and development using blade coating, e.g. OPV module production. 

\section{Material and Methods}\label{}

\subsection{Theoretical description of wet film thickness for trapezoidal applicator geometry}
The theory of wet film formation dates back to the work of Landau-Levich \cite{Landau_Levich_1988} and Derjaguin \cite{Derjaguin_1993}. In their work, they describe how the wet film is generated on a flat plate which is dragged out of a liquid (mixture of solvent and solute) reservoir. They assume a relatively fast coating velocity, so effects of the drying process can be neglected (Landau-Levich regime). Both articles propose an analytical expression for the generated wet film

\begin{equation}
    h = 1.34 \cdot R \cdot Ca^{2/3},
    \label{eqh}
\end{equation}

where $h$ denotes the wet film thickness (SI-unit: $m$) and $R$ the radius of the down-stream meniscus (SI-unit: $m$). The capillary number $Ca$ is defined as 

\begin{equation}
    Ca = \frac{\mu \cdot u}{\sigma},
    \label{eqCa}
\end{equation}

where $\mu$ is the viscosity of the liquid (SI-unit: $Pa$ $s$), $u$ is the velocity of the plate (SI-unit: $m/s$) and $\sigma$ is the surface tension of the liquid in air environment (SI-unit: $N/m$). The theory was applied to meniscus-guided coating processes where cylindrical applicator geometries are used to predict the resulting wet film thickness \cite{Nickel2012, Park_Han_2009}.\\
It is found that the liquid volume beneath the applicator affects the radius of the down-stream meniscus and consequently also the resulting wet film thickness. Moreover, for cylindrical applicators also the contact angle between the applicator and the coating liquid varies with the volume, which makes the system even more complex. In contrast, applicators with trapezoidal-shaped tips provide constant contact angles between applicator and liquid as long as the volume is greater than a critical minimal value. In \cite{Gutenev2003}, a tilted plate moves along a substrate. The gap between plate and substrate is filled with liquid. The authors derive an expression for the radius from fundamental geometrical considerations, if $h/R<<1$

\begin{equation}
    R(S) = \sqrt{\frac{S}{\tan(\frac{\pi - \varphi}{2}) + \frac{\varphi - \pi}{2}}}.
    \label{eqR}
\end{equation}

\begin{figure}[H]
    \centering
    \includegraphics[width=0.9\textwidth]{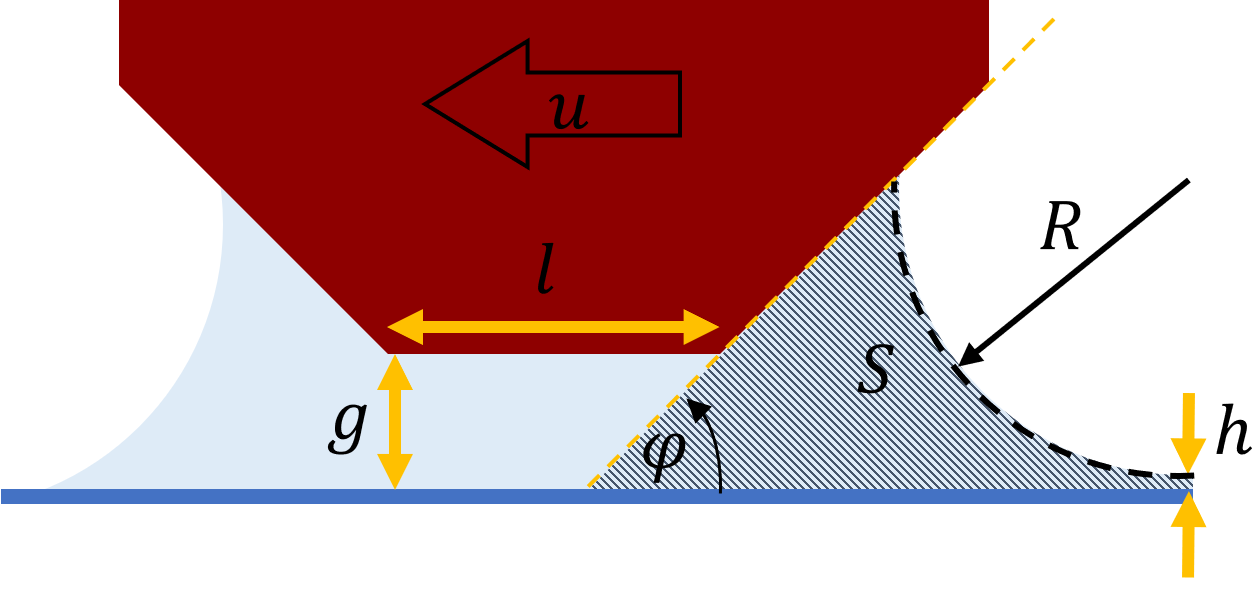}
    \caption{Cross-sectional view of the meniscus-guided coating process. In the process, the coating liquid (light blue) is injected between the applicator tip (red) and the substrate (dark blue). The down-stream meniscus $R$, which mainly determines the wet film thickness, is indicated and is a function of the area $S$ (hatched area).}
    \label{figCrossArea}
\end{figure}

Here, $S$ denotes the cross-sectional area of the liquid (SI-unit: $m^2$) and $\varphi$ the opening angle in radians between the substrate and the plate. In Figure \ref{figCrossArea}, the cross-sectional view of the coating process is shown and all relevant parameters for the calculation of the down-stream meniscus radius are indicated, apart from the applicator width $w$ (SI-unit $m$) that is orthogonal to the drawing plane.

The applicator moves with a velocity $u$ at a height of $g$ over the substrate (SI-unit: $m$). Together with the length of the applicator tip $l$ (SI-unit: $m$), a minimal volume can be defined. The angle $\varphi$ can be derived from the applicator geometry. Assuming a symmetrical liquid distribution beneath the applicator, which is valid for small velocities, the relevant liquid area $S$ (SI-unit: $m^2$) can be derived from basic geometrical considerations


\begin{equation}
    S(V_0) = \frac{ \frac{g^2}{\tan(\varphi)} + \frac{V_0}{w} - g \cdot l}{2}.
    \label{eqS}
\end{equation}

Inserting equations \ref{eqCa}, \ref{eqR}, and \ref{eqS} in \ref{eqh}, an analytical expression to predict the wet film thickness as a function of the coating velocity $u$ and the initial volume $V_0$ can be derived

\begin{equation}
    h_{theo}(u,V_0) = 0.95 \sqrt{\frac{\left[ \frac{g^2}{\tan(\varphi)} + \frac{V_0}{w} - g \cdot l \right]}{\tan(\frac{\pi - \varphi}{2}) + \frac{\varphi - \pi}{2}}} \left( \frac{\mu \cdot u}{\sigma} \right) ^{2/3}.
    \label{eqTheo}
\end{equation}

However, this theoretical expression is only able to predict layer thicknesses for small coating distances where the volume loss is neglectable. 

\subsection{The accelerated coating process for a constant film thickness}
For OPV module-relevant coating distances (i.e. $>$10 $cm$), the initial volume $V_0$ will decrease considerably by the process itself. For a coating distance $x$, the fluid volume beneath the applicator $V(x)$ can be written as

\begin{equation}
    V(x) = V_0 - h \cdot w \cdot x.
\end{equation}

In a previous work \cite{Gumpert_Janßen_Brabec_Egelhaaf_Lohbreier_Distler_2023}, we presented a computational fluid dynamics (CFD) simulation as a digital twin of the coating process which only requires a few, experimentally easily accessible, liquid properties to predict the wet film thickness with good accuracy. Based on that, we derived a linear fit function for one specific OPV material system to predict the wet film thickness

\begin{equation}
    h_{fit}(u, V_0) = (b + c \cdot V_0)\cdot u^{2/3},
    \label{eqFit}
\end{equation}

where $b$ and $c$ are fit parameters which can be determined by the simulation results. \\
In both, Eq. \ref{eqTheo} and Eq. \ref{eqFit}, it can be seen that the wet film thickness is a function of the fluid volume and the velocity. Therefore, the loss of liquid during the coating process can be compensated by gradually increasing the velocity of the applicator during the coating process, a technique that can easily be implemented.
For a constant film thickness $h_{c}$, an expression $u_{theo}(x)$ to describe the velocity at position $x$ can be derived from Equation \ref{eqTheo} 

\begin{equation}
    u_{theo}(x) = 1.08 \frac{\sigma}{\mu} \left\{h_{c}  \sqrt{\frac{\tan(\frac{\pi - \varphi}{2}) + \frac{\varphi - \pi}{2}}{\left[ \frac{g^2}{\tan(\varphi)} + \frac{V_0}{w} - h \cdot x - g \cdot l \right]}}\right\}^{3/2},
    \label{equTheo}
\end{equation}

and a respective expression $u_{fit}(x)$ can be derived from Equation \ref{eqFit}

\begin{equation}
    u_{fit}(x) = \left[ \frac{h_{c}}{b + c (V_0 - h \cdot w \cdot x)} \right]^{3/2}.
    \label{equFit}
\end{equation}

\subsection{Material system P3HT:O-IDTBR}
Different experiments are carried out to validate the proposed accelerated coating approach. In all of these experiments, the material system P3HT:O-IDTBR is used which is based on the photoactive material combination of the donor polymer poly(3-
hexylthiophene) (P3HT) and the non-fullerene acceptor
(5Z,5’Z)-5,5’-((7,7’-(4,4,9,9-tetraoctyl-4, 9-dihydros-
indaceno[1,2-b:5,6-b’]dithiophene-2,7- diyl)bis(benzo
[c][1,2,5]thiadiazole-7,4-diyl))bis(methanylylidene bis
(3-ethyl-2-thioxothiazolidin-4-one))) (O-IDTBR). The material system offers good processing and stability properties \cite{Strohm2018, Armin2021} and thus, has a high relevance for large-area industrial OPV applications. In our previous work \cite{Gumpert_Janßen_Brabec_Egelhaaf_Lohbreier_Distler_2023}, we experimentally determined the different fluid-related properties of P3HT:O-IDTBR (dynamic viscosity $\mu$, surface tension $\sigma$, and the contact angles with the applicator $\alpha$ and the glass substrate $\beta$) which are required for both, the numerical simulation and the theoretical approach. Further information about the material system can be found in our previous paper.

\section{Results and Discussion}\label{}

\subsection{Comparison of theory, simulation and experiment}
\begin{figure}[H]
    \centering
    \includegraphics[width=0.9\textwidth]{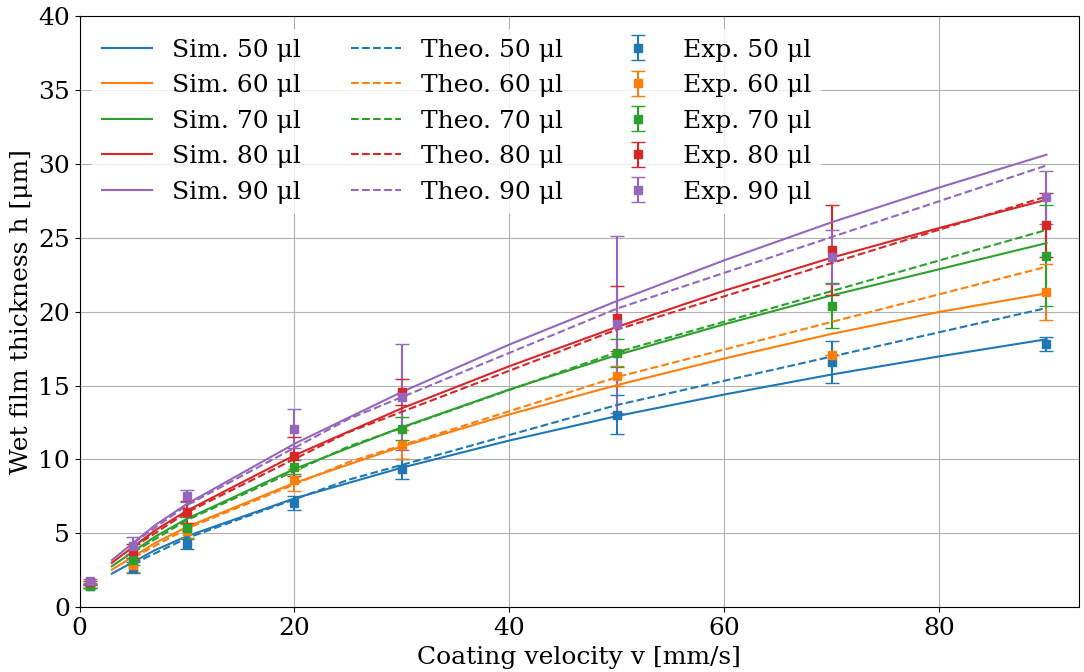}
    \caption{Theoretically predicted (dashed lines), simulated (solid lines) and experimentally measured (squares with corresponding error bars) wet film thicknesses as a function of the coating velocity for different initial liquid volumes.}
    \label{fighTheoSimExp}
\end{figure}

In 
\cite{Gumpert_Janßen_Brabec_Egelhaaf_Lohbreier_Distler_2023}, experiments are performed to determine the wet film thickness of a blade-coated P3HT:O-IDTBR layer as a function of $V_0$ and $u$. The developed CFD simulation is shown to reproduce the experimental results with high precision. In Figure \ref{fighTheoSimExp}, we compare these experimental and simulated results with calculations from our newly derived analytical expression (Eq. \ref{eqTheo}).

\begin{figure}[H]
    \centering
    \includegraphics[width=0.6\textwidth]{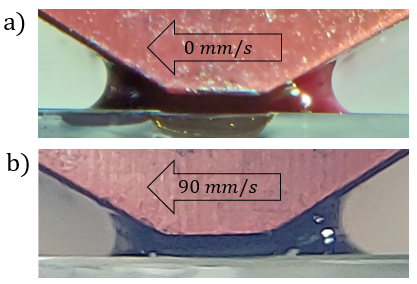}
    \caption{Cross-sectional view of the experimental setup, where the tip of the applicator (red) and the coating ink (black) are visible. a) Symmetric fluid distribution, if the applicator is at rest (0 $mm/s$). b) Asymmetric fluid distribution, if the applicator moves with high coating velocity (90 $mm/s$).}
    \label{figCrossSecExp}
\end{figure}

Overall, the theoretical prediction shows great agreement with the simulation and experimental results. For velocities below 50 $mm/s$, a maximal difference between simulated and theoretically predicted wet film thickness of $\sim$ 0.5 $\mu m$ is obtained. For faster coating velocities, the theoretically calculated wet film thicknesses slightly differ from simulation and experimental results. For such high velocities, the assumption of a symmetrically distributed liquid volume beneath the applicator (as shown in Figure \ref{figCrossArea}) has been shown not to be fulfilled anymore \cite{Gumpert_Janßen_Brabec_Egelhaaf_Lohbreier_Distler_2023}. The meniscus is dragged more to the down-stream side, which can explain the differences in wet film thickness obtained by the different investigation methods. Photographs from the experiment confirm this explanation (see Figure \ref{figCrossSecExp}).

If the applicator is at rest (Figure \ref{figCrossSecExp} a)), the coating fluid is symmetrically distributed beneath the applicator, whereas a unsymmetrical fluid distribution can be observed for a coating velocity of 90 $mm/s$ (Figure \ref{figCrossSecExp} b)). The fluid distribution in the CFD simulations can be seen in the supplementary data.\\
The results clearly confirm the validity of our theoretical model, which has been specially adapted to the geometry and physics of our system of investigation.

\subsection{Influence of fluid parameters on the wet film thickness}
\begin{figure}[H]
    \centering
    \includegraphics[width=0.8\textwidth]{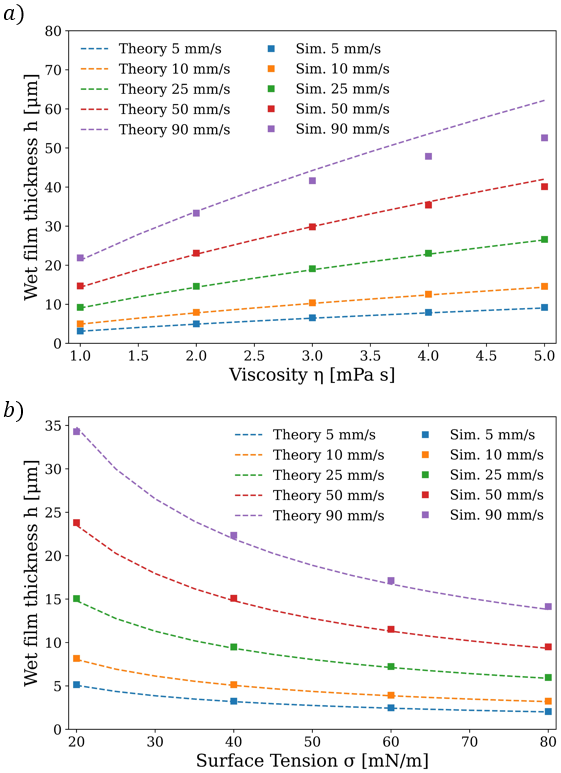}
    \caption{The wet film thickness is plotted as a function of viscosity $\mu$ (a)) and surface tension $\sigma$ (b)) for an initial liquid volume $V_0$ of 80 $\mu l$ for different coating speeds $u$. In both a) and b), square markers indicate the simulation results and the theoretical calculations are plotted as dashed lines.}
    \label{figSurfViscoSweep}
\end{figure}

In \cite{Gumpert_Janßen_Brabec_Egelhaaf_Lohbreier_Distler_2023}, all simulations and experiments are performed with the same coating fluid, namely a P3HT:O-IDTBR ink, which is an established material for the active layer in OPV. For this material, the surface tension and viscosity have been determined experimentally, since these crucial fluid parameters for the coating process and the respective simulation. However, the development of novel materials for OPV is a current and very active research topic \cite{Zhang2022}. The creation of a large variety and number of new materials aims at further pushing the power conversion efficiency \cite{Zhan2022, Zhu2022} or to improve the stability of the active material \cite{Xian_Zhang_Xu_Liu_Zhou_Peng_Li_Zhao_Chen_Fei_et, Liu_Zhang_Li_Li_Huang_Jing_Cheng_Xiao_Wang_Han_et}. However, each new material provides different fluid properties impacting the wet film thickness.\\
Therefore, our proposed simulation/theory approach is tested with respect to its flexibility regarding variation of fluid properties in the following. In Figure \ref{figSurfViscoSweep}, the theoretical and simulated results for an initial volume $V_0$ of 80 $\mu l$ are shown. The wet film thicknesses are determined for different viscosities (Figure \ref{figSurfViscoSweep} a)) and surface tensions of the fluid (Figure \ref{figSurfViscoSweep} b)) by numerical simulations and the analytical expression.

In Figure \ref{figSurfViscoSweep} a), results from the CFD simulation and the theory for velocities are very consistent for velocities below 50 $mm/s$. For faster coating velocities, the theory predicts slightly higher thicknesses compared to the simulation. This can again be explained by an asymmetrical fluid distribution under the applicator. For the dependence on the surface tension of the fluid, numerical simulation and analytical expression show an excellent agreement for all values of $\mu$ and $\sigma$ (see Figure \ref{figSurfViscoSweep} b)).\\
Both, our CFD simulation model as well as our adapted theoretical function, are shown to provide the same results regarding variations in fluid properties, which strongly suggests that these results also match the real experimental findings, and thus both be used to predict and manipulate the coating thickness, as will be shown in the following chapter.\\
In the supplementary information, a similar figure to Fig. \ref{figSurfViscoSweep} can be found for an initial volume $V_0$ of 70 $\mu l$.

\subsection{Homogeneous wet films processed by accelerated coating}
For the investigation and implementation of an accelerated coating strategy, we used a Zehntner ZAA 2300 blade coating device that was electrically modified by Automatic Research GmbH to enable the pre-programming of a time-dependent speed profile. The velocity of the applicator $u$ can be changed every 100 $ms$. As a consequence, we can apply our proposed fluid-specific acceleration strategy to compensate for the steadily decreasing film thickness upon coating due to the constant loss of fluid volume. Thus, the steadily decreasing film thickness could be compensated by using a fluid-specific accerlation profile.
To apply the compensation strategy, the previous equations needs to be reformulated to describe the velocity as a function of time. Based on Equation \ref{equTheo}, the velocity as a function of time can be written as




\begin{equation}
    u_{theo}(t) = 1.08 \frac{\sigma}{\mu} \left\{h_{c} \sqrt{\frac{\tan(\frac{\pi - \varphi}{2}) + \frac{\varphi - \pi}{2}} {\left[ \left(\frac{g^2}{\tan(\varphi)} + \frac{V_0}{w} - g \cdot l\right)^{7/4} - 1.9 \frac{t \cdot \sigma \cdot h_{c}^{5/2} \left(\tan\left(\frac{\pi - \varphi}{2}\right) + \frac{\varphi - \pi}{2}\right)^{3/4}}{\mu } \right]^{4/7}} }\right\}^{3/2} .
    \label{equttheo}
\end{equation}

An almost equivalent, but more compact, equation can be derived from the fit function approach

\begin{equation}
    u_{fit}(t) = \left\{ \frac{h_{c}}{[(c \cdot V_0 + b)^{5/2} - 2.5 \cdot c \cdot h_{c}^{5/2}\cdot w \cdot t]^{2/5}} \right\}^{3/2}.
    \label{equtfit}
\end{equation}

With these equations, the necessary velocities for a coating process with constant thickness $h_{c}$ can be calculated and the respective time-velocity tables can be programmed in the device. For three different initial volumes $V_0$ (70, 80, and 90 $\mu l$) and a targeted constant wet film thickness $h_{c}$ of 10 $\mu m$ the time-velocity tables are calculated with the fully theoretical- (Eq. \ref{equttheo}) and the fit function- (Eq. \ref{equtfit}) approach for 15 $cm$. The resulting velocity profiles are shown in the following figure as a function of the corresponding substrate position $x$.

\begin{figure}[H]
    \centering
    \includegraphics[width=0.7\textwidth]{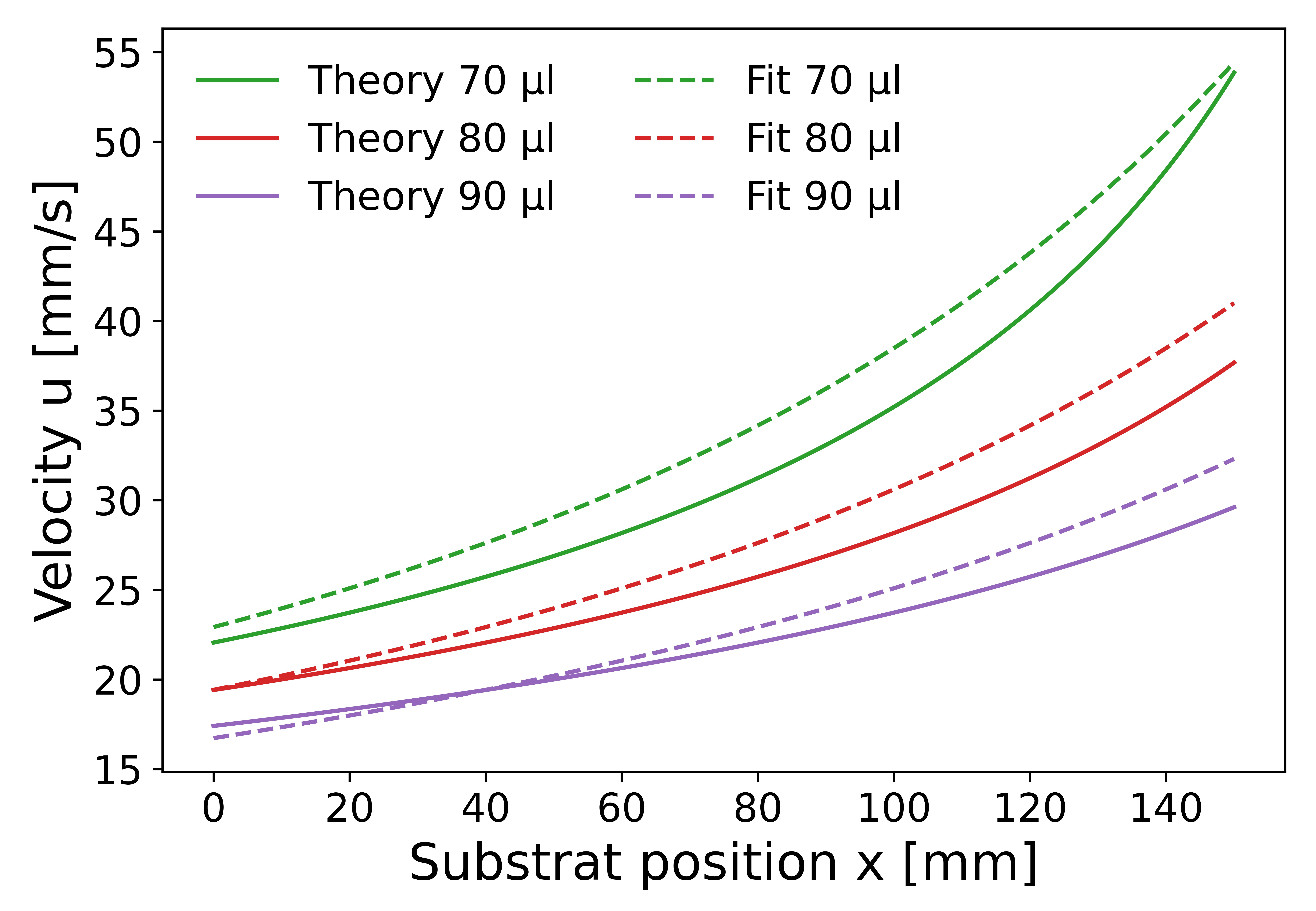}
    \caption{The coating velocity $u$ as a function of the substrate position is determined by the theoretical expression (solid lines) and the fit function-approach (dashed lines) for three different $V_0$ (70, 80, and 90 $\mu l$).}
    \label{figUs}
\end{figure}

\begin{figure}[H]
    \centering
    \includegraphics[width=0.9\textwidth]{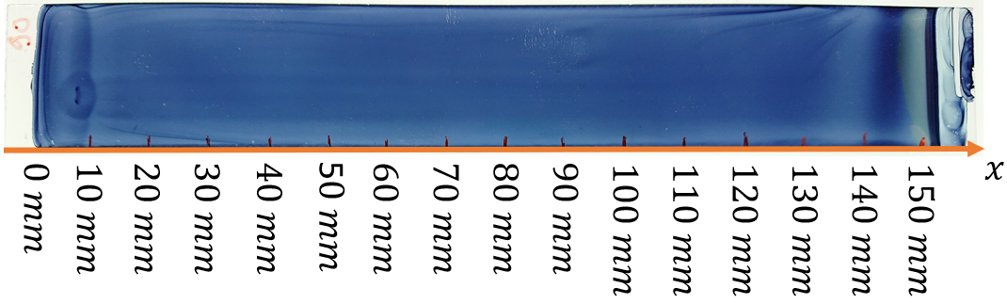}
    \caption{Photograph of a P3HT:O-IDTBR film deposited with accelerated balde coating using 90 $\mu l$ initial volume. The coating direction $x$ is indicated as well as the regarding distance from the coating start point ($x = 0 mm$).}
    \label{figConstSample}
\end{figure}

For all initial volumes $V_0$, the velocities profiles, based on theoretical considerations and on the fit function, show some differences. The reason for the discrepancies can be seen in Figure \ref{fighTheoSimExp}. Each approach predicts a slightly different velocity to manufacture a 10 $\mu m$ thick wet film e.g. for an initial volume of 90 $\mu l$. \\
For the experimental validation of the accelerated coating strategy, the coating velocities for three different initial volumes (70, 80, and 90 $\mu l$) are calculated with Eq. \ref{equtfit}. In Figure \ref{figConstSample}, a photograph of a substrate after the drying process is shown. The accelerated coating strategy was applied to create a uniform film thickness of 10 $\mu m$ onto the substrate, the initial volume of the coating ink was 90 $\mu l$. 

\begin{figure}[H]
    \centering
    \includegraphics[width=0.7\textwidth]{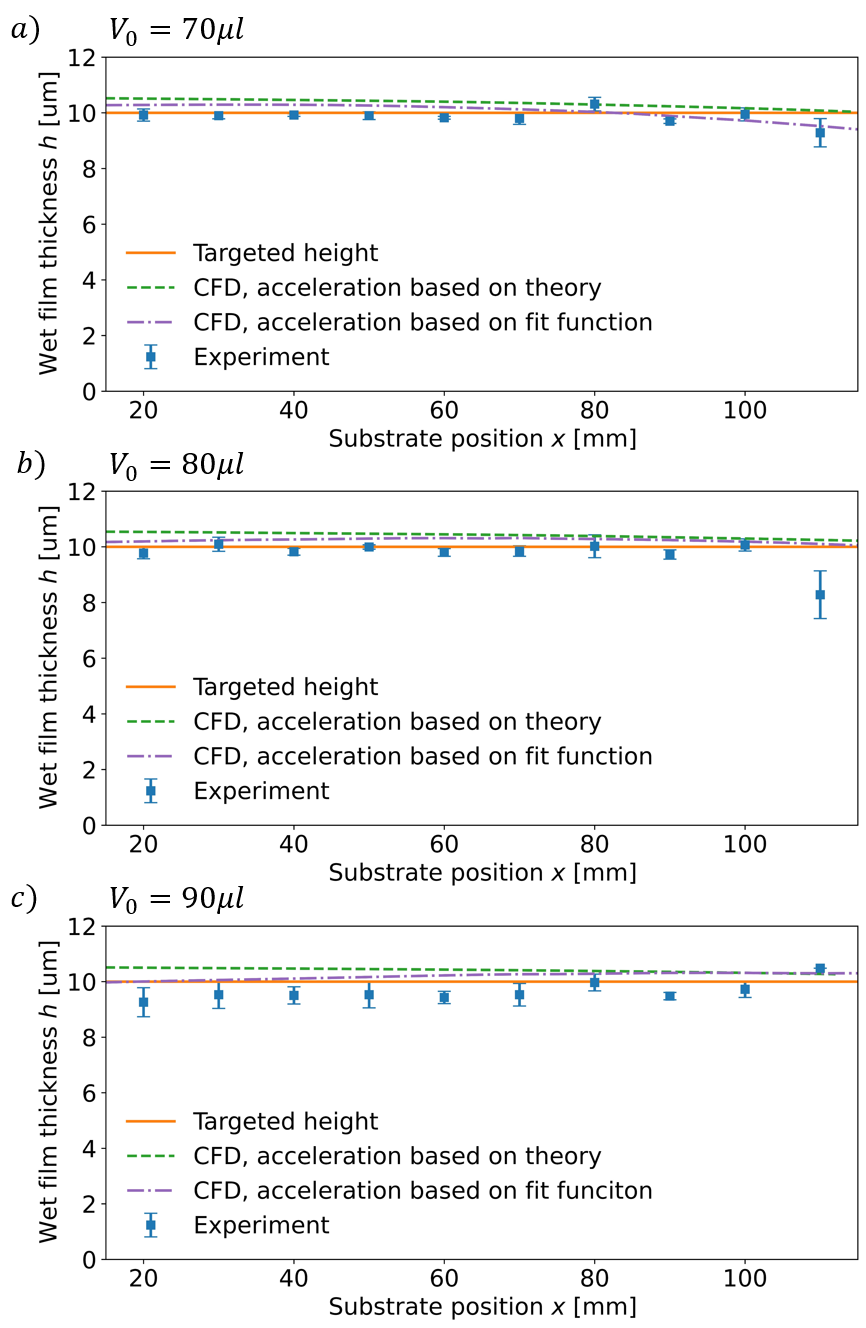}
    \caption{a) - c) The experimentally determined wet film thicknesses (blue squares with error bars, averaged over two samples per variation) are shown for the initial volumes 70 $\mu l$, 80 $\mu l$, and 90 $\mu l$, respectively. CFD results, which are based on the acceleration strategy according to Eq. \ref{equttheo} (green dashed lines) and Eq. \ref{equtfit} (purple dash-dotted lines). The targeted wet film thickness (10 $\mu m$) is plotted as solid orange line.}
    \label{figConsth}
\end{figure}

The dry film thicknesses are measured to validate the accelerated coating strategy. From the composition of the material system, we can conclude the wet film thickness from the dry film thickness \cite{Gumpert_Janßen_Brabec_Egelhaaf_Lohbreier_Distler_2023}. The starting point of the measurements is 20 $mm$ to exclude effects of the initial movement of the applicator on the film thickness. The thickness of the coating is measured every 10 $mm$ until a coating distance of 110 $mm$ has been reached. The respective calculated wet film thicknesses are plotted in Figure \ref{figConsth} a) - c) (blue squares and error bars). CFD simulations were conducted to compare the two methods of calculating the velocity profiles. CFD results, where the velocity profile is calculated according to the theoretically derived equation are plotted as green dashed lines whereas purple dash-dotted lines indicate the CFD results, where the velocity profile is based on the fluid-specific fit function.

\begin{figure}[H]
    \centering
    \includegraphics[width=0.7\textwidth]{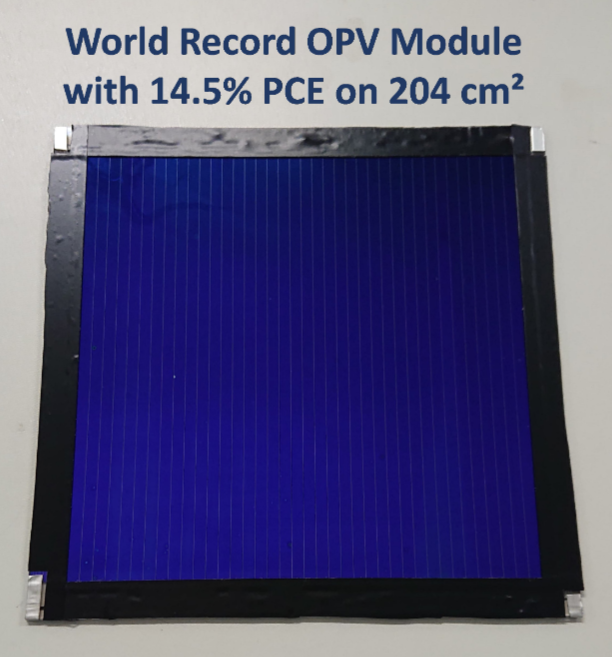}
    \caption{Photograph of the organic photovoltaic (OPV) module with a new world record power conversion efficiency (PCE) of 14.5 $\%$ on 204 $cm^2$ enabled by accelerated blade coating. The substrate size is 165 mm x 165 mm.}
    \label{figOPVmodule}
\end{figure}

For all investigated initial volumes, the measured film thickness is constant over the whole evaluable distance and very close to the targeted value (orange line). In addition, the results of the corresponding CFD simulations, which also incorporate the respective acceleration profiles, are plotted as dashed line, and show as well an excellent agreement with the targeted and experimentally obtained values. Note: The deviation of the data point at 110 mm for an initial volume of 80 $\mu l$ is due to an experimental error caused by an inhomogeneous drying process close to the end position of the applicator bar, which also explains the relatively large error bar.\\
This data proves that both our simulative model and our analytical equation can be used to create a full set of process parameters, which can be easily applied in practice to fabricate homogeneous coating over large distances.\\
In fact, we already implemented this approach very recently into the manufacturing process of high-performance large-area OPV modules, which enabled us to yield a new world record efficiency for OPV modules with 14.5 $\%$ on 200 $cm^2$ \cite{NREL2024, FAU2024}. This module is depicted in Figure \ref{figOPVmodule}  and demonstrates the excellent film homogeneity over the whole module area, which is crucial for its high power conversion efficiency.

CFD simulations are performed where the applicator acceleration is determined once with the theoretical expression and once with the formula, which is based on the fit function. The results of the CFD simulations are plotted in Figure \ref{figConsth} a) - c) as green dashed and purple dash-dotted lines for the theoretical- and fit function-approaches, respectively. Both approaches result in almost identical simulated wet film thicknesses. According to the CFD simulation, the acceleration, which is based on theoretical considerations, leads to slightly thicker wet films compared to the acceleration, according to Equation \ref{equtfit}.  

\section{Conclusion}\label{}

In this article, we derived a theoretical expression to predict the height of the deposited wet film in the doctor blading process. The predicted wet film thicknesses of the theoretical description have been compared with experimental data and results of a corresponding CFD simulation and show great agreement in all investigated cases, which includes variations of the ink’s viscosity and surface tension. Thus, both the proposed theoretical description and the simulation model enable an extremely time- and resource-efficient approach for high-throughput research, which is crucial for fields like OPV, where a large number of novel materials is newly developed in a very short time.\\
Expressions have been derived for both the theoretical equation and the fit function to describe an acceleration of the applicator to compensate the consumption of fluid during the coating process and to consequently achieve uniform layers with predefined thickness. This acceleration strategy was experimentally tested for long coating distances and the results of experiment, simulation, and calculation all show similar values close to the targeted thickness over the whole coating distance.\\
This newly developed approach was finally used to fabricate uniform large-area coatings for high-performance organic solar modules and allowed us to break the world record for OPV module efficiency, which underlines the importance and applicability of this work.

\section*{CRediT authorship contribution statement}
\textbf{Fabian Gumpert:} Investigation, Methodology, Visualization, Writing - Original Draft. \textbf{Annika Janßen:} Investigation, Validation, Writing - Review $\&$ Editing. \textbf{Robin Basu:} Validation, Writing - Review $\&$ Editing. \textbf{Christoph J. Brabec:} Resources, Writing - Review $\&$ Editing. \textbf{Hans-Joachim Egelhaaf:} Resources , Writing - Review $\&$ Editing. \textbf{Jan Lohbreier:} Supervision, Writing - Review $\&$ Editing. \textbf{Andreas Distler:} Supervision, Writing - Review $\&$ Editing.

\section*{Declaration of competing interest}
The authors declare no competing financial interest or personal relationship that could have appeared to influence the work reported in this paper.

\section*{Data availability}
Data will be made available on request.

\section*{Acknowledgements}
Funding: This work was supported by the Bavarian State Ministry for Science and Art via the Energy Campus Nürnberg (EnCN); the European Union’s Horizon 2020 research and innovation program [Grant number 952911 and 101007084].



  \bibliographystyle{elsarticle-harv} 
  \bibliography{biblio.bib}

\end{document}